\documentclass{article}

\usepackage{xspace}

\newcommand{\method}{DSA\xspace}
\usepackage{graphicx}
\usepackage{algorithm}
\usepackage{algorithmicx}
\usepackage{algpseudocode}
\usepackage{amsmath}
\usepackage{amsthm}
\newtheorem{theorem}{Theorem}
\usepackage{caption}
\usepackage{tikz-cd}%
\usepackage[final]{neurips/neurips_2023}
\usepackage{accents}
\usepackage{bbm}

\def\*#1{\mathbf{#1}}

\newcommand{\x}{\mathbf{x}}

\newcommand{\Tr}{\text{Tr}}

\newcommand{\fR}{\mathbb{R}}

\newcommand*{\dt}[1]{\accentset{\raisebox{1pt}{\scalebox{0.4}{$\bullet$}}}{#1}}

\usepackage{titlesec}

\usepackage[utf8]{inputenc} % allow utf-8 input
\usepackage[T1]{fontenc}    % use 8-bit T1 fonts
\usepackage{hyperref}       % hyperlinks
\usepackage{url}            % simple URL typesetting
\usepackage{booktabs}       % professional-quality tables
\usepackage{amsfonts}       % blackboard math symbols
\usepackage{nicefrac}       % compact symbols for 1/2, etc.
\usepackage{microtype}      % microtypography
\usepackage{xcolor}         % colors

\title{Beyond Geometry: Comparing the Temporal Structure of Computation in Neural Circuits with Dynamical Similarity Analysis}
\author{{\large \bf Mitchell Ostrow}, {\large \bf Adam Eisen}, {\large \bf Leo Kozachkov}, {\large \bf Ila Fiete} \\
  Department of Brain and Cognitive Sciences, MIT \\
  Cambridge, MA, USA \\ 
  \texttt{\{ostrow,eisenaj,leokoz8,fiete\}@mit.edu}}

\begin{document}

\maketitle

\begin{abstract}
% \documentclass{article}
% \usepackage{setspace}
% \doublespacing

How can we tell whether two neural networks utilize the same internal processes for a particular computation? This question is pertinent for multiple subfields of neuroscience and machine learning, including neuroAI, mechanistic interpretability, and brain-machine interfaces. Standard approaches for comparing neural networks focus on the spatial geometry of latent states. Yet in recurrent networks, computations are implemented at the level of dynamics, and two networks performing the same computation with equivalent dynamics need not exhibit the same geometry. To bridge this gap, we introduce a novel similarity metric that compares two systems at the level of their dynamics, called Dynamical Similarity Analysis (\method). Our method incorporates two components: Using recent advances in data-driven dynamical systems theory, we learn a high-dimensional linear system that accurately captures core features of the original nonlinear dynamics. Next, we compare different systems passed through this embedding using a novel extension of Procrustes Analysis that accounts for how vector fields change under orthogonal transformation. In four case studies, we demonstrate that our method disentangles conjugate and non-conjugate recurrent neural networks (RNNs), while geometric methods fall short. We additionally show that our method can distinguish learning rules in an unsupervised manner. Our method opens the door to comparative analyses of the essential temporal structure of computation in neural circuits. 

\end{abstract}

\section{Introduction}

Comparing neural responses between different systems or contexts plays many crucial roles in neuroscience. These include comparing a model to experimental data (as a measure of model quality, \citet{schrimpf_brain-score_2018}), determining invariant states of a neural circuit (\cite{chaudhuri_intrinsic_2019}), identifying whether two individual brains are performing a computation in the same way, aligning recordings across days for a brain-machine interface (\cite{degenhart_stabilization_2020}), and comparing two models (e.g. to probe the similarity of two solutions to a problem, \citet{Pagan2022.11.28.518207}). Current similarity methods include $R^2$ via Linear Regression, Representational Similarity Analysis, Singular Vector Canonical Correlation Analysis, Centered Kernel Alignment, and Procrustes Analysis (\cite{schrimpf_brain-score_2018,kriegeskorte_representational_2008,raghu2017svcca,williams_generalized_2022,duong2022representational}). Crucially, these all compare the \textit{geometry} of states in the latent space of a neural network. 

Identifying shared neural computations between systems calls for methods that compare similarity at the level of the fundamental computation. In the brain, computations are instantiated by the emergent dynamical properties of neural circuits, such as fixed points, invariant and fixed point manifolds, limit cycles, and transitions between them (\cite{amari_competition_1977,hopfield_neurons_1984,hopfield_computing_1986,seung_how_1996, zhang_representation_1996,hahnloser_digital_2000, burak_accurate_2009,Wang2002-co,sussillo_generating_2009,churchland2012neural,chaudhuri_intrinsic_2019,vyas2020computation,khona_attractor_2021}.Geometric similarity metrics fall short in two manners when they are applied to dynamical systems. First, response geometries can differ due to sampling differences (\cite{chaudhuri_intrinsic_2019}) or other nonfundamental causes even though the dynamics are topologically equivalent. In such cases, measures of state-space geometry do not fully capture the core similarities between two systems, a false negative situation. Conversely, systems may exhibit distinct dynamics over similar geometries in the state space (\cite{galgali_residual_2023}). In this case, geometric measures may fail to distinguish two distinct systems, a false positive situation. Therefore, we suggest that dynamical similarity should be the preferred lens through which to compare neural systems. 

In dynamical systems, a key notion of similarity is called topological conjugacy: the existence of a homeomorphism that maps flows of one system onto flows of the other.  When two systems are conjugate, they have the same qualitative behavior. Given two dynamical systems $f: X \rightarrow X$ and $g: Y \rightarrow Y$ with invertible mapping $\phi : X \rightarrow Y$, conjugacy is defined as:
\begin{equation}
g \circ \phi = \phi \circ f \label{eqn_conj}
\end{equation}
Here, we develop a data-driven method called Dynamical Similarity Analysis (\method ), which combines work from two previously-unrelated fields, machine learning for dynamical systems and statistical shape analysis. In brief, \method returns a similarity metric describing how two systems compare at the level of their dynamics. First, \method uses the insight that appropriate high-dimensional embeddings can be used to describe a highly nonlinear system in a way that is globally linear. The features that enable this transformation are referred to as Koopman Modes (\cite{mezic2005spectral}) and can be identified via the Dynamic Mode Decomposition (DMD) (\cite{snyder2021koopman,brunton_modern_2021}). Subsequently, \method employs a statistical shape analysis metric to compare vector fields within the Koopman Mode embeddings, thereby assessing the similarity between the systems' dynamics. This shape analysis measures how far a pair of systems are from being homeomorphic. Unlike most spatial shape analyses, which require averaging over multiple noisy trials in the same condition to arrive at one vector for each condition-time point (but see \cite{duong2022representational}), small amounts of noise create a rich representation of neural dynamics in the estimated dynamics matrix. The importance of flows in response to intrinsic (natural) or induced perturbations in capturing dynamics has been highlighted in \citet{seung_how_1996,yoon_specific_2013,chaudhuri_intrinsic_2019,galgali_residual_2023}, as their evolution over time can uniquely identify dynamic structure. 

Our results demonstrate that \method effectively identifies when two neural networks have equivalent dynamics, whereas standard geometric methods fall short. \method therefore has great relevance for neuroscience, providing another method to validate the fit of a model to the neurophysiology data. It additionally opens the door to novel data-driven analyses of time-varying neural computation. We expect our method to be interesting to computational neuroscientists due to its theoretical richness, valuable to deep learning researchers as a theoretically backed and rigorously justified tool through which to interpret and compare neural networks, and useful for experimental neuroscientists due to its simplicity of implementation and ease of interpretation. %due to its linearity.
\paragraph{Contributions} We develop a general method, \method, for comparing two dynamical systems at the level of their dynamics. To do so, we introduce a modified form of Procrustes Analysis that accounts for how vector fields change under orthogonal transformations. We apply our method in four test cases, each demonstrating novel capabilities in neural data analysis: \textbf{(a)} Our method can identify dynamical similarities underlying systems with different geometries, which shape metrics cannot. \textbf{(b)} Conversely, \method can distinguish systems with different dynamics despite similar geometries, which shape metrics cannot. \textbf{(c)} We show that our method empirically identifies topological similarities and differences between dynamical systems, as demonstrated by invariance under geometric deformation and variance under topological transformation. \textbf{(d)} We demonstrate how to use \method to disentangle different learning rules in neural networks without supervision, by comparing how representations change across training. \footnote{https://github.com/mitchellostrow/DSA/}

%Finally, we will provide open source, GPU-accelerated code for our method.

\section{Methods}
\subsection{Theoretical Background: Koopman Operators and Dynamic Mode Decomposition (DMD)}

A common approach to modeling nonlinear dynamics is to approximate them with linear systems, as those are more tractable and interpretable. One standard method involves finding fixed points of the dynamics, and then locally linearizing the system around each fixed point, to determine whether infinitesimal changes diverge from or converge to the fixed point. Another is to describe nonlinear dynamics via locally linear models, a separate one for each point in state space. These methods do not generate a description of the global dynamics. The former method also requires injecting perturbations, which may not be possible for experimental systems and pre-collected datasets. 

The field of dynamical systems theory suggests a different approach, on which a new class of data-driven methods are based. Specifically, Koopman operators theoretically embed nonlinear systems into infinite-dimensional Hilbert space, which permits an exact and globally linear description of the dynamics (\cite{koopman1931hamiltonian}). Practical applications use the Dynamic Mode Decomposition (DMD) (\cite{schmid_dynamic_2010}) to create a finite-dimensional embedding that approximates the Koopman operator (\cite{brunton_extracting_2016,brunton_chaos_2017,snyder2021koopman,dubois_data-driven_2020}). The DMD identifies a linear transition operator for the dynamics:  $X(t+\Delta t) = AX(t)$, where $X$ comprises a data matrix (consisting of functions of the observed states of the dynamical system). When applied to a sufficiently rich $X$, the DMD can capture global dynamical structure. A key challenge of the method is in constructing an appropriate space for $X$ which is closed under the operation of $A$ (a Koopman invariant subspace, \cite{brunton2016koopman}). We employ the Hankel Alternative View of Koopman (HAVOK, \cite{brunton_chaos_2017,arbabi_ergodic_2017}), which constructs $X$ as a function of the delay-embedding of the observed variables (a Hankel Matrix), such that one column of $X = [ -x(t)-  -x(t - \tau)- ... -x(t - p\tau)- ]^T$, where $p$ is the number of included delays and $\tau$ is the time-step between each measurement.  The delay-embedding approach eliminates the difficulty of learning a nonlinear transformation of the data, and improves estimation capabilities over partially-observed systems (\cite{kamb_time_delay_2020}). The latter capability arises from Takens' delay embedding theorem, which states that a delay embedding is diffeomorphic to the original system--that is, a partially-observed system can be fully reconstructed using a sufficient number of delays (\cite{takens}). However, our method could be used with a range of embedding algorithms in place of delay embeddings. 
% While Takens' theorem was introduced for deterministic settings and noise complicates the task of state-space reconstruction (\cite{casdagli1991}), the delay embedding method is an efficient way of embedding a system to improve linear predictivity. 

\subsection{Dynamical Similarity Analysis (\method)}

Now, we introduce {\em Dynamical Similarity Analysis}, a general method that leverages advances in both Statistical Shape Analysis and Dynamical Systems Theory to compare two neural systems at the level of their dynamics. Simply put, we use Koopman operator theory to generate a globally linear embedding of a nonlinear system, via delay embeddings as in HAVOK, then use DMD to define the linear transition operators in this space. Finally, we apply a novel extension of the Procrustes Analysis shape metric to the resultant DMDs from two systems. This set of steps constitutes \method. 

Suppose we have two dynamical systems defined by the following equations:

\begin{equation}
    \dt{\mathbf{x}} = f(\mathbf{x},t) \quad \mathbf{x} \in \mathbb{R}^n \hspace{1cm} \dt{\mathbf{y}} = g(\mathbf{y},t) \quad \mathbf{y} \in \mathbb{R}^m
\end{equation}

We sample data from each system $X \in \mathbb{R}^{c \times k_1 \times t_1 \times n}$ and $Y \in \mathbb{R}^{c \times k_2 \times t_2 \times m}$. Here, $c$ indicates the number of conditions observed, which in neuroscience and deep learning refer to different input stimuli or contexts. $k_i$ indicates the number of trials per condition, and $t_i$ indicates the number of time steps observed per trial.  Note that the observations must be sampled uniformly in time (\cite{takens}). Next, we produce delay-embedded Hankel tensors with a lag of $p$: $H_x \in \mathbb{R}^{c \times k_1 \times (t-p-1) \times np}$ and likewise for $H_y$. Using the Hankel Alternative View of Koopman (HAVOK) approach (\cite{brunton_chaos_2017}), we flatten all but the last dimension of $H$ and fit reduced-rank regression models with rank $r$ to the eigen-time-delay coordinates of the data, where the target is the coordinates of the next time step ($V_{x,r}'$):
\begin{equation}
    V_{x,r}' = A_x V_{x,r} \mbox{  where  } H_x^T = U_x \Sigma_x V_x^T \mbox{  and } V_{x,r} = V_x[:,1:r]
\end{equation}
Only the rank of the HAVOK models explicitly need to be the same so that the two DMD matrices can be compared. Note that the eigen-time-delay coordinate formulation is equivalent to PCA whitening (Supplementary Information Section \ref{supp:whiten}), which \citet{williams_generalized_2022} highlights to be important. The predictive capability of our model (via MSE, $R^2$, AIC, or others) can be used to identify the optimal rank $r$ and number of delays $p$. In our experiments, hyperparameters for the DMD were chosen to ensure both sufficient capability to fit the dynamics in question as well as tractable computation times. Fitting the HAVOK model to each system returns our DMD matrices: $\*A_x, \*A_y \in \mathbb{R}^{r \times r}$, which we compare using the modified Procrustes Analysis algorithm detailed next.

\begin{figure}[ht!]
    \centering
    \includegraphics[width=\linewidth]{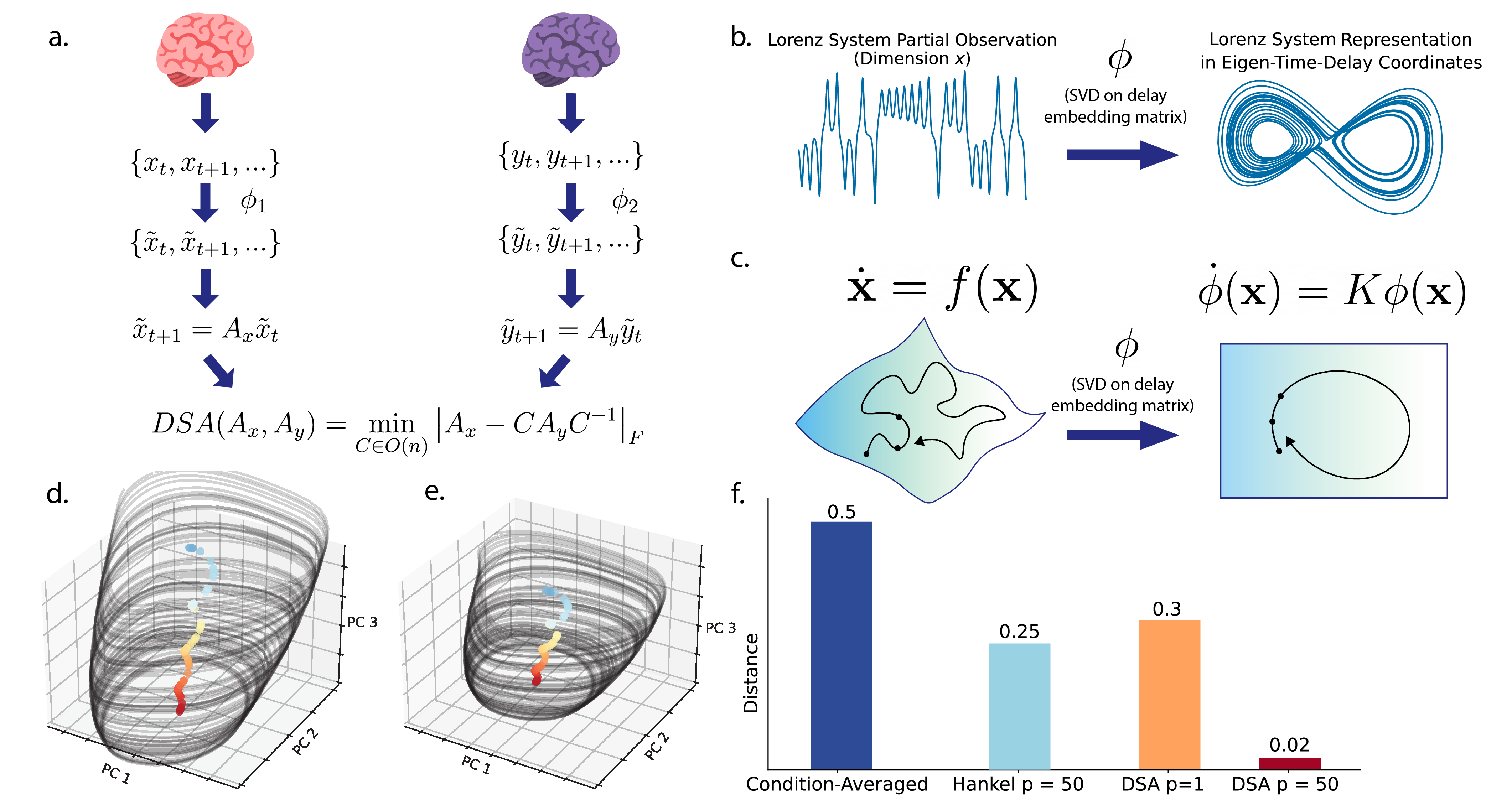}%{neurips/Methods_Fig1_v2.png}
     \caption{\textbf{Schematic and example of \method }. \textbf{a.} Schematic of the DSA pipeline. 
\textbf{b.} Schematic of attractor reconstruction with delay embeddings. \textbf{c.} Schematic of the Koopman Operator linearization over time-delay coordinates. \textbf{d,e.} PCA trajectories (black lines) of a GRU (left) and Vanilla RNN (right) trained to produce sine waves of a range of frequencies (\cite{sussillo_opening_2013}). Colored dots indicate center points of the limit cycles, colored by frequency of the output wave. \textbf{f.} Comparison of measured dissimilarity between the networks in \textbf{d,e} with (from left to right) Procrustes on the condition-averaged trajectories, Procrustes on the delay-embedded condition-averaged trajectories, \method  on the original data, and \method  on the delay embedding. Lower is more similar. Each metric uses angular distance, which ranges from 0 to $\pi$.}
    \label{fig:schematic}
\end{figure}

\paragraph{Procrustes Analysis over Vector Fields} Procrustes Analysis is a shape metric which compares the geometric similarity of two datasets after identifying an optimal orthogonal transformation to align them (\cite{statshapes}). To use it in \method, we must modify the notion of orthogonal transformations to be compatible with linear systems. Recall that the Procrustes metric solves the following minimization problem (\cite{williams_generalized_2022}):
\begin{equation}
    \textnormal{d}(\*X,\*Y) = \min_{\*C \in O(n)}||\*X - \*C \*Y ||_F 
\end{equation}
Where $O(n)$ is the orthogonal group ($C^TC = I$), $F$ indicates the Frobenius norm, and $X,Y$ are two data matrices. A distance metric that is invariant to orthogonal transformations captures the intuitive notion that there is no privileged basis across neural networks. 

However, vector fields, represented by the DMD matrices $\*A_x,\*A_y$ above, transform differently than vectors X and Y (see Supplementary Information Section \ref{supp:orthog} for an illustration). The invariance we seek is that of a conjugacy, or change of basis, between linear systems, which is reflected in the similarity transform, $CA_xC^{-1}$. Note that this is a special case of equation \ref{eqn_conj} over linear systems. Thus we introduce a novel similarity metric, which we call {\em Procrustes Analysis over Vector Fields}:  
\begin{equation}\label{dsanormal}
\textnormal{d}(\*A_x,\*A_y) = \min_{\*C \in O(n)}||\*A_x - \*C \*A_y \*C^{-1}||_F 
\end{equation}
Here, $C$ is once again an orthogonal matrix. The angular form of this distance can be written as:
\begin{equation}\label{dsaangular}
\textnormal{d}(\*A_x,\*A_y) = \min_{\*C \in O(n)} \textnormal{arccos} \frac{\*A_x \cdot (\*C \*A_y \*C^{-1})}{\left | \*A_x \right |_F \left | \*A_y \right |_F } 
\end{equation}
where  $\*A_x \cdot (\*C \*A_y \*C^{-1})$ is computed as $ \Tr(\*A_x^T\*C\*A_y\*C^{-1})$.
In the Supplementary Information Section \ref{equivalency}, we prove an equivalence between \method and the 2-Wasserstein distance between the two DMD matrices' eigenspectrum, in special cases. This connection highlights the  grounding of \method in Koopman operator theory to measure conjugacy: when two dynamical systems are conjugate, their Koopman eigenspectra are equivalent (Proof in Supplementary Information \ref{supp:conj}, \cite{budisic_applied_2012,meziclinearization}). We derive equation \ref{dsanormal} from first principles and prove that it is indeed a proper metric in the Supplementary Information (\ref{supp: metric}); the proper metric property is crucial for various downstream machine learning pipelines such as clustering or classification (\cite{williams_generalized_2022}).  

Equations \ref{dsanormal} and \ref{dsaangular} are nonconvex optimization problems similar to those in \citet{jimenez_fast_2013}, although we solve them with gradient-based optimization. Across our experiments, we found that the metric asymptotes in roughly 200 iterations using the Adam optimizer (\cite{kingma2014adam}) with a learning rate of 0.01. To optimize over the orthogonal group, $O(n)$, we utilize an extension of the Cayley Map, which parameterizes $SO(n)$ (Supplementary Information \ref{supp:optimization}). In our experiments, we compare the response of \method with direct application of Procrustes to the data matrices as it is the most analogous shape metric. If only the identification of conjugacy is desired, metric properties can be relaxed by optimizing $C$ over the larger group of invertible matrices, $GL(n)$. % although this is a much more challenging problem as $GL(n)$ is noncompact.

%Note that other similarity metrics could potentially be modified to fit the \method algorithm.

\section{Results}
To demonstrate the viability of our method, we identified four illustrative cases that disentangle geometry and dynamics, and additionally performed an ablation analysis of \method. First, a set of neural networks that are different by shape analyses but are dynamically very similar; second, a set of neural networks that are similar by shape analyses but are dynamically are very different; third, a ring attractor network whose geometry we can smoothly and substantially deform while preserving the attractor topology; finally, a line attractor network whose topology is transformed into a ring attractor by adding periodic boundary conditions.

\subsection{Ablation Study}
Importantly, both the reduced-rank regression and delay-embedding in \method are required to isolate dynamical structure. We demonstrate this in Fig. \ref{fig:schematic}f, where we compare the hidden state trajectories of an RNN and GRU solving the Sine Wave task (\cite{sussillo_opening_2013}) using the standard Procrustes Analysis (via the \href{https://github.com/ahwillia/netrep}{netrep} package, \cite{williams_generalized_2022}), delay-embedded Procrustes Analysis (an ablation), \method with no delay embedding (another ablation), and \method score. To be exhaustive, we also computed a similarity metric that compares purely temporal information, defined as the angular Procrustes distance over the power spectrum of the RNN's hidden state trajectories. On the RNNs in Fig. \ref{fig:schematic}, this metric reported 1.41.  Only \method identifies the systems as highly similar, indicating that both the delay embedding and the DMD are necessary to abstract geometric particulars from each system.

\subsection{\method captures dynamical similarity between RNNs despite differences in geometry.} 

We trained a collection of 50-unit RNNs on the well-known 3-bit Flip Flop task from \citet{sussillo_opening_2013}. In this task, an RNN is presented with a time series from three independent channels, each sporadically occurring pulses of either -1 or 1. The network's objective is to continuously output the last pulse value for each channel until it is updated by a new pulse (hence the Flip-Flop name). Following \citet{maheswaranathan_universality_2019}, We varied architecture (LSTM, GRU, UGRNN, Vanilla RNN), activation function (ReLU, Tanh), learning rate (.01,0.005), and multiple different seeds across each set of hyperparameters, ultimately collecting a dataset of 240 networks. We only saved a network if it solved the task with MSE less than 0.05. After training, we tested all models on the same 64 input sequences and extracted the recurrent states, upon which we apply Procrustes and \method in parallel to compare networks. 

\begin{figure}[ht!]
    \centering
    \includegraphics[width=\linewidth]{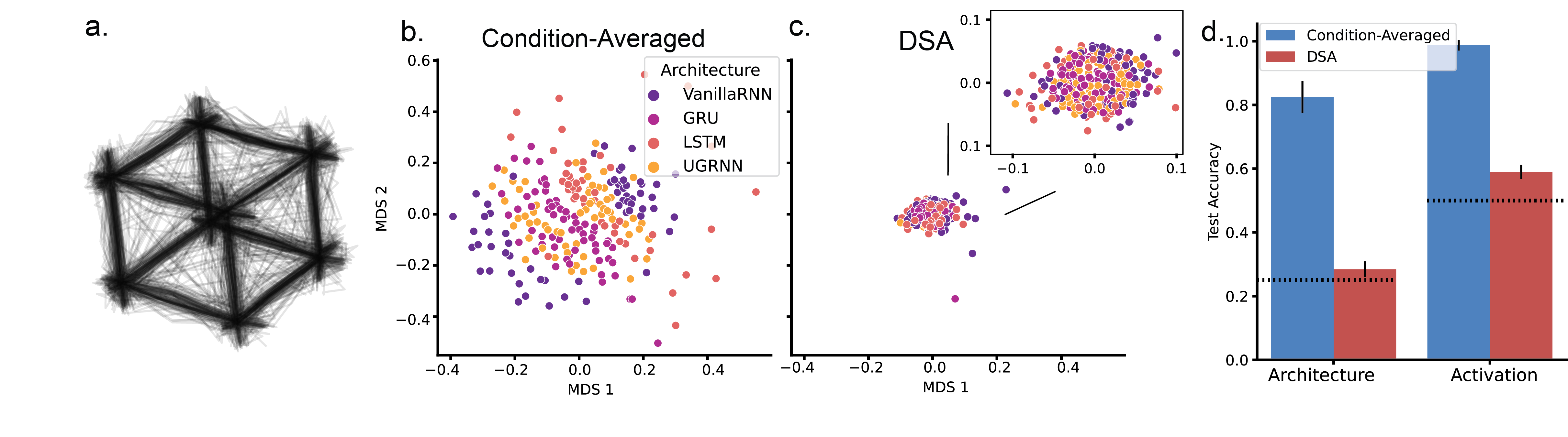}
    \caption{\textbf{Same dynamics, different shape, only identified as similar with \method }. Dots indicate a single trained network. Analysis of RNNs trained on the 3-bit FlipFlop task. \textbf{a.} PCA trajectories of a single trained RNN. \textbf{b.} MDS projection of the dissimilarity matrix (DM) computed across condition-averaged hidden states between each RNN (architecture indicated by color). In this view of the data, RNNs cluster by architecture. \textbf{c.} MDS embedding of the DM generated from \method of the same set of RNNs as in b. Here, RNNs do not cluster by architecture. \textbf{d.} Classification test accuracy of condition-averaged Procrustes and \method similarity spaces on both architecture and activation labels. Dotted lines indicate chance.}
    \label{flipflop}
\end{figure}

Our RNNs have different geometry but equivalent attractor topology (\cite{maheswaranathan_universality_2019}). Fig.\ref{flipflop}a. displays the trajectories of individual trials in a sample network in the first three principal components, which capture on average 95.9\% of the total variance. The computational solution to this task is characterized by eight stable fixed points at the vertices of a cube that correspond to the eight possible unique memory states ($2^3$). Importantly, this cubic structure is preserved across all networks, even though particular geometric aspects of the trajectories may differ. Standard shape analyses differentiate architectures and activation function by geometry, which we replicated in Fig. \ref{flipflop}b (architecture) and in the Supplementary Information (activation, \ref{supp:mds_act}). 

We fit individual HAVOK models with 75 delays and rank 100 to sampled trajectories from each network. For computational efficiency, we reduced the dimensionality of the network to 10 using Principal Component Analysis (PCA) before applying HAVOK. In Supplementary Information \ref{supp:heatmap}, we demonstrate that HAVOK's predictivity is robust to hyperparameter variation. Because the variance explained from the later PCs is negligible, they can be disregarded while preserving the original dynamics. This improves HAVOK's fit to the data and reduces the memory and time complexity of DSA. For each pair of networks, we compared these states directly using Procrustes Analysis and on the DMD matrices using Procrustes Analysis on Vector Fields. This yields two dissimilarity matrices (DMs), which we reduced to two dimensions using Multidimensional Scaling (MDS, \cite{kruskal_multidimensional_1964}). We plot the results in this similarity space in Fig. \ref{flipflop}, where each point is a trained network. As expected, the networks cluster by architecture (Fig. \ref{flipflop}b) and activation function (Supplementary Information) when Procrustes is applied, indicating that aspects of their geometry differ. However, the networks intermingle under \method (Fig. \ref{flipflop}c, inset zoomed in), and have pairwise dissimilarities close to zero, indicating that their underlying dynamics are equivalent, which is correct (\cite{maheswaranathan_universality_2019}). We quantified these differences by applying a linear SVM to the dissimilarity matrices and plotted the test accuracy in Fig. \ref{flipflop}d. This further underscores the differences between Procrustes and \method in Fig. \ref{flipflop}b and c, as \method's accuracy is close to chance, whereas Procrustes achieves high accuracy on both labels.
 
\subsection{\method identifies dynamical differences despite geometric similarity}

\begin{figure}
    \centering
    \includegraphics[width=\linewidth]{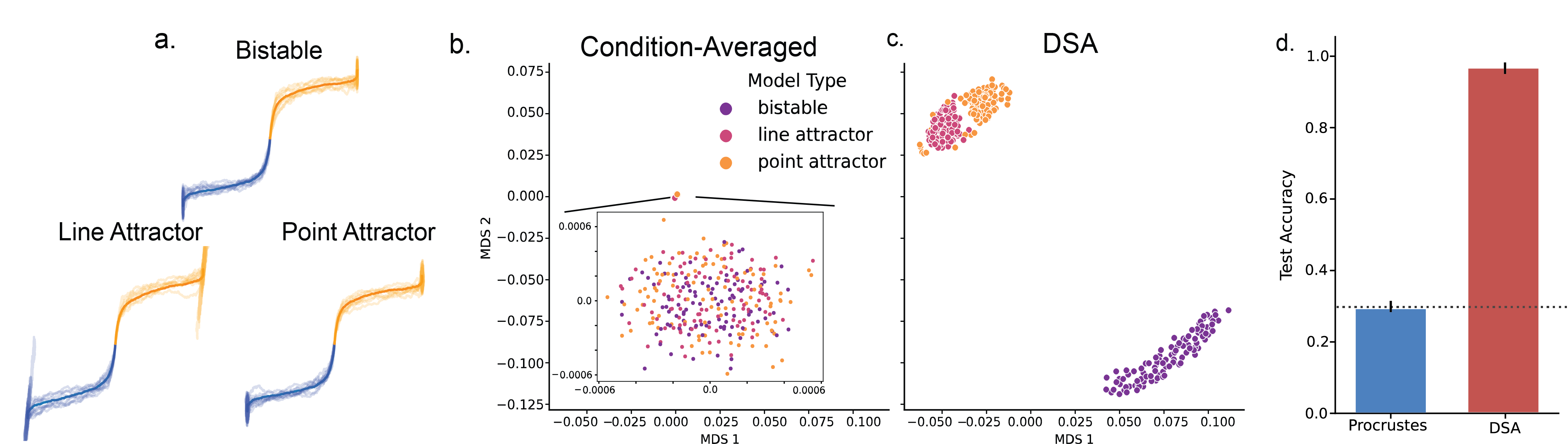}%{neurips/neurips3_v2.png}
    \caption{\textbf{Different dynamics, same shape, only distinguished by \method.} \textbf{a.} Sample single-trial (faded) and condition-averaged trajectories of the three systems in two dimensions (x and y axes). Trials start at the center point and move away in a direction depending on the condition (colored). \textbf{b.} Procrustes-MDS on the dissimilarity matrices of the three network types. \textbf{c.} \method-MDS of the three networks. \textbf{d. }Classification test accuracy of condition-averaged Procrustes and \method similarity spaces. Dotted line indicates chance.}
    \label{galgali}
\end{figure}

Next, we sought to display the converse, namely that our similarity method can identify differences in dynamical structure even when spatial analysis cannot. To do so, we examined a set of three noisy recurrent networks from \citet{galgali_residual_2023}. These systems implement a bistable pair of attractors, a line attractor, and a single stable fixed point at the origin. These systems describe three strategies in a classical perceptual decision-making task from systems neuroscience: unstable evidence integration, stable integration, and leaky integration. Their dynamical equations are included in the Supplementary Information. Here, the inputs to the latter two networks were adversarially optimized so that the condition-averaged trajectories are indistinguishable from the first network (Fig. \ref{galgali}a). These networks have intrinsic noise in their hidden states, which enables \method to distinguish systems beyond condition averages. We simulated 200 noisy trajectories from 100 networks per class, each with randomly sampled parameters.

To confirm that the condition-averaged trajectories are indistinguishable, we visualize the condition-averaged and single-trial trajectories in Fig. \ref{galgali}a for each system. We fit HAVOK models with 100 lags and a rank of 50 for each network and computed DSA and Procrustes pairwise (5,000 total comparisons). As before, we plot a low-dimensional MDS projection of these dissimilarity matrices (\ref{galgali}). When the two types of similarities are compared side by side (spatial similarity, Fig. \ref{galgali}b, and \method, Fig. \ref{galgali}c), it is evident that only our method is capable of easily distinguishing dynamic types from the data alone. This is quantified by almost perfect test accuracy via a linear classifier in Fig. \ref{galgali}d. Thus, \method can efficiently extract features that describe the underlying dynamical system from data alone, which better reflect the nature of the computation that is being performed in RNNs.

\subsection{\method is invariant under geometric deformations that preserve attractor topology}

\begin{figure}
    \centering
    \includegraphics[scale=0.5]{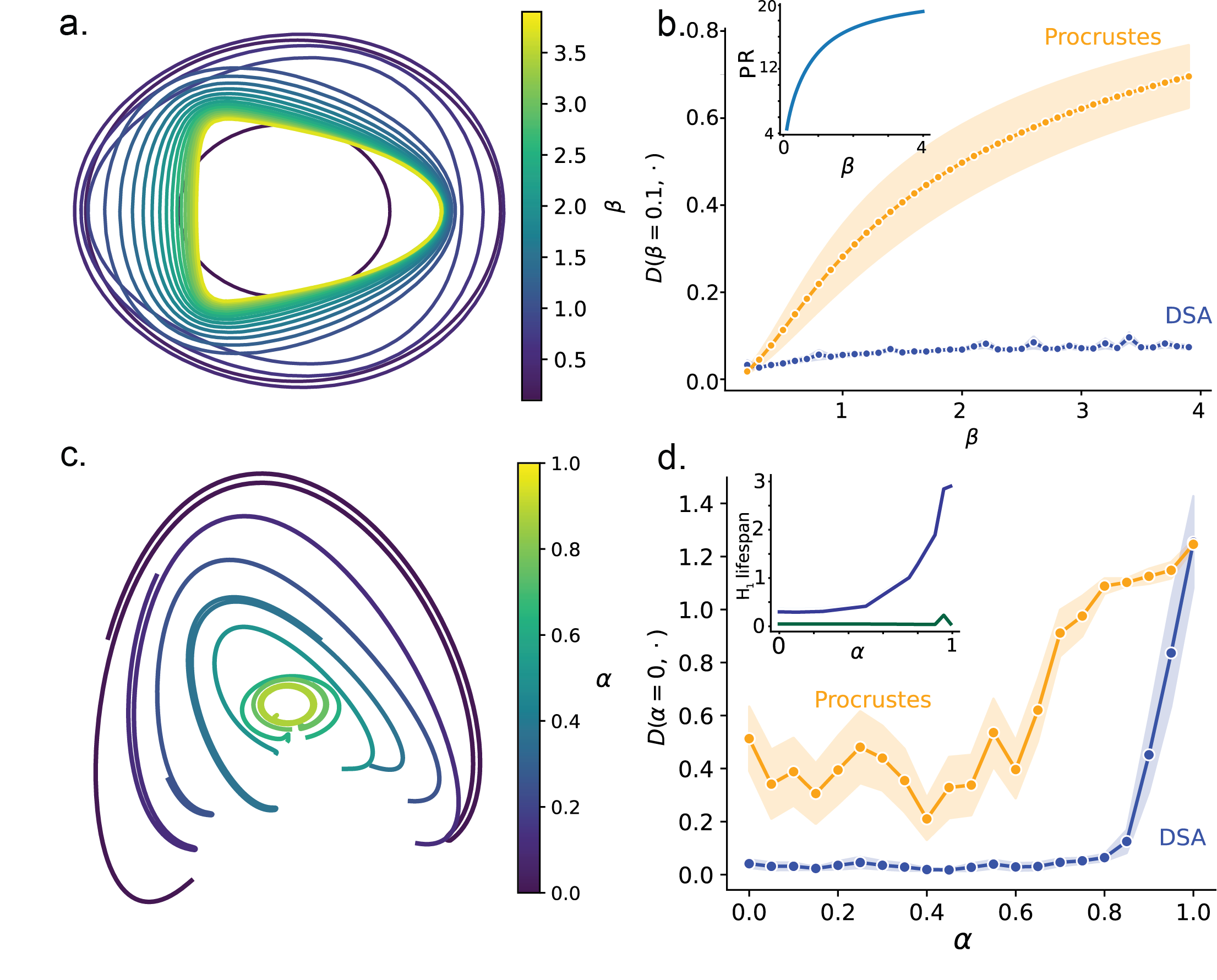}%{neurips/fig4_wtopologyv2.png}
    \caption{\textbf{\method is invariant under geometric deformation of a ring attractor, and sensitive to the transformation of a ring attractor into a line. }Top row displays scaling of $\beta$ in the sigmoid, making the ring progressively less planar. The bottom row displays topological transformation of a ring into a line via $\alpha$. The network is a curve line attractor at $\alpha < 1$, and is a ring attractor at $\alpha = 1$. \textbf{a, c}. Trajectories along the ring plotted in the top two PCs of the network, colored by magnitude of the deformation parameter. In \textbf{c}, trajectories are scaled for visualization purposes. \textbf{b,d}. Distance (D) to $\beta=0.1$, and $\alpha=0.0$ in both Procrustes and \method. Importantly, the topology only changes in the last row, at $\alpha=1.0$.  Inset \textbf{b}: Participation Ratio of the ring, a natural continuous measure of dimensionality based on explained variance ratios of PCA eigenvalues (\cite{gaoganguli}). Inset \textbf{d}. Persistent Homology Analysis of the ring: top 2 lifespans of Betti number 1 features, indicating that the data becomes more ring-like as $\alpha$ increases.}
    %\textbf{b,f}. Visualization of the bump on the ring network as the deformations are applied, colored by magnitude of the deformation. \textbf{c,g}. Explained Variance Ratio (EVR) of the top two PCs as the deformations are applied on the x-axis.
    %\textbf{b}. Explained Variance Ratio (EVR) of the top two PCs progressively decreases as the ring deforms, indicating that the ring becomes less planar \textbf{
    % \textbf{e.} Persistent Homology analysis at each $\alpha$ value. At each value, there is a significant cycle, demonstrated by a large $H_1$ lifespan.
    \label{ringnetwork}
\end{figure}

Next, we examined how \method responds when smooth geometric deformations are applied to a dynamical system. Here, we chose to study a ring attractor network, a system known to track head direction in the brain, and whose topology is a ring embedded in $n$ neurons (\cite{zhang_representation_1996,chaudhuri_intrinsic_2019,skaggs_model_1995}). The neural activity on the ring is instantiated as a bump of local activity, which is stable due to the localized and translationally-symmetric connectivity profile. In each trial, we randomly sampled $n$ uniformly between 100 and 250, and drove the network with constant input and dynamical noise, leading the bump to rotate around the ring at a roughly constant velocity (we used \href{https://github.com/louiskang-group/wang-2022-multiple-bumps}{code} from \cite{wang2022}). The network's dynamics are detailed in the Supplementary Information (\ref{supp:ring}). Importantly, after simulation, the saved activations can be transformed by a continuous deformation that preserves the ring topology while changing the geometry: 
% \vspace{-2mm}
\begin{equation}
    r = \frac{1}{1+\exp{(-\beta s)}} 
\end{equation}
The variable $s$ describes the synaptic activations of the ring network (the dynamical variable), and $r$ describes the neural activations, which we compare with \method and Procrustes. Here, $\beta$ controls the width of the bump. As $\beta$ increases, the width of the bump becomes progressively tighter, which makes the ring less planar (see Fig. 1 in \citet{kriegeskorte_neural_2021} for a depiction). We scaled the magnitude of $\beta$ from 0.1 to 4. In the inset of Fig. \ref{ringnetwork}b, we quantified the ring's dimensionality across $\beta$, measured as the participation ratio of the principal components' eigenvalues: $PR = \frac{(\sum_i \lambda_i)^2}{\sum_i \lambda_i^2}$ (\cite{gaoganguli}). 
 
 After applying these transformations to simulated trajectories from our network, we applied \method and Procrustes to respectively compare all $\beta$s to the lowest $\beta$  in our dataset. In Fig. \ref{ringnetwork}a we visualize the hidden state trajectory in the first two PCs as the ring is progressively transformed with $\beta$. We fit HAVOK models with a lag of 10 and a rank of 1000 to trajectories at each level of deformation. When we compared the network using \method and Procrustes, we found that only \method is invariant across these transformations (Fig. \ref{ringnetwork}b). While the value for \method is not strictly zero, this may be due to approximation or numerical error in HAVOK and the similarity metric, as it remains close to zero throughout (between 0.05 and 0.12). This suggests that \method identifies similarity at the level of the topology of a dynamical system. 

%This can also be measured by calculating the cumulative explained variance ratio of the first two PC's, which starts at ~99\% at small $\beta$ but drops to ~30\% at large $\beta$ (Supplementary Information). 
% Yet the dissimilarity measured by \method is quite stable, with a standard deviation of 0.0175 across all trials \textbf{and} all $\beta$ values. On the other hand, Procrustes had a standard deviation of 0.13 across trials \textbf{per} deformation value.

\subsection{\method responds to changes in topology}

Finally, we assessed the converse to the previous analysis: does \method respond to changes in topology? Using the same ring attractor network, we continuously varied the strength of its periodic boundary conditions to convert the ring into a curved line segment attractor by ablating some neurons from the network on one end of the ring. Our ring is specified by a cosine kernel that defines how each neuron connects to its neighbors. This kernel has a length parameter $l$ which specifies how far apart two neurons can connect. To completely eliminate the boundary conditions, $l$ neurons must therefore be ablated. Using an ablation length $c$ as the number of neurons to ablate in the ring, we defined the following parameter to normalize the transformation across different network sizes.
\begin{equation}
    \alpha = 1 - \frac{c}{l}
\end{equation}
When $\alpha = 1$, there is no ablation and the network is perfectly a ring. When $\alpha < 1$, the network breaks into a line segment attractor. As $\alpha \rightarrow 0$ the ring becomes progressively less curved. To ensure that network sizes were consistent across all values of $\alpha$, we initialized the network size to be $n + c$ before ablation, where $n$ is the desired final number of neurons. We simulated networks of sizes between 200 and 300 neurons. Instead of a constant drive of input as before, we drove the network with a fixed magnitude input $u$ whose sign flipped stochastically. We chose this input to prevent the bump of activity from getting trapped on one end of the line attractors.

When we compared each value of $\alpha$ of the network to $\alpha=0$ with \method and Procrustes, we identified that \method reports values close to 0 until $\alpha$ becomes close to 1.0, when it jumps almost to $\pi/2$. We plotted these results in Fig. \ref{ringnetwork}, where panel \textbf{c} depicts the low-dimensional visualization of the line and the ring, and \textbf{d} displays our metric's response. Note that because the data are noisy, we expected \method to jump slightly before 1.0, which we empirically verified to be around $\alpha=0.9$ in our simulations. Here we used only 10 delays (for a total embedding size between 2000 and 3000) and rank 100, but found that our results generalized to much larger DMD models as well. This further solidifies the notion that \method captures topological similarity of two dynamical systems.
\begin{figure}[h]
    \centering
    \includegraphics[width=\linewidth]{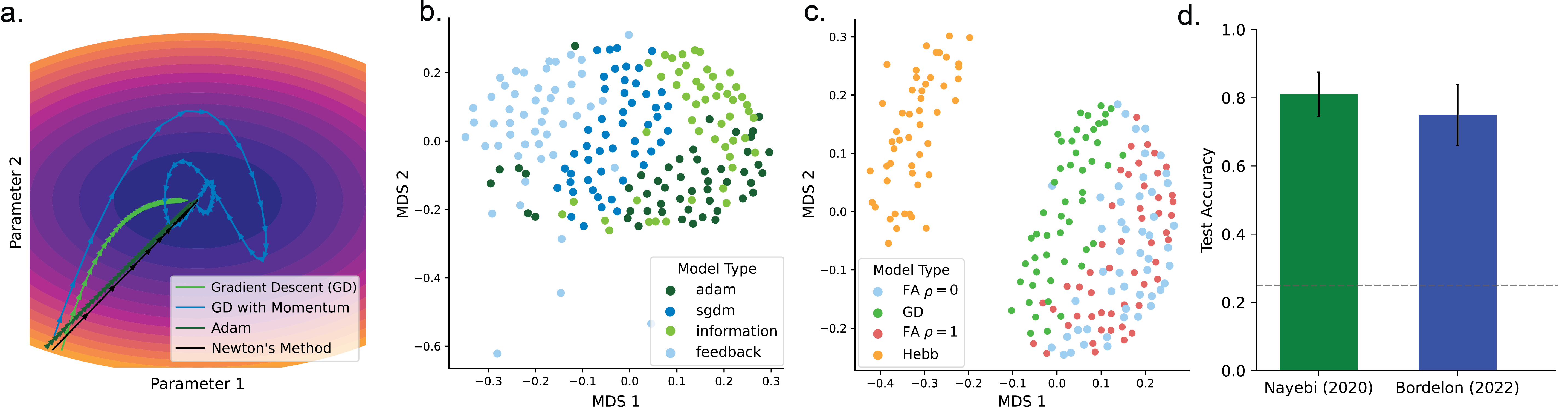}%{neurips/neurips5.png}
    \caption{\textbf{\method unsupervisedly disentangles learning rules.} \textbf{a.} Schematic example of different algorithms minimizing $3x^2 + 2y^2$. \textbf{b.} \method MDS plot of hidden activation observables from networks trained on ImageNet \textbf{c.} \method MDS plot of the dynamics of learning in 3-layer linear networks. \textbf{d.} Test accuracy of a linear decoder on the \method similarity space for the model classes. Bars indicate standard deviation over a wide range of \method hyperparameters.}
    \label{fig:learning_rules}
     \vspace{-6mm}
\end{figure}
\subsection{\method can disentangle learning rules in an unsupervised fashion}\label{section:learning_rules}

It is still unclear what learning rules the brain uses. Recent progress includes the design of novel biologically-plausible learning rules (\cite{lillicrap_random_2016,hinton2022forwardforward}), experiments and modeling work that suggests signs of gradient-based learning in the brain (\cite{sadtler_neural_2014,Humphreys2022.12.08.519453,payeur_neural_2023}), supervised methods to classify learning rules from aggregate statistics in neural data (\cite{nayebi_identifying_2020}), and theory describing how representations evolve across training in different learning rules (\cite{bordelon_influence_2022}). Because artificial networks are initialized with random seeds, their trajectories in the activity space across learning will almost always have different geometry, despite the fact that they may converge to the same minimum. However, because the trajectories arise from the same dynamical system, \method should be able to capture similarities between learning trajectories from different initializations. We visualize a variety of learning rule trajectories in Fig. \ref{fig:learning_rules}a to demonstrate how qualitatively different their dynamics can be. We applied \method on two test cases from the literature to demonstrate that we can empirically disentangle artificial networks trained with different learning rules. To do so, we fit \method to trajectories of the neural representations (or observables) across training epochs. 

In the first case, we drew 21 aggregate statistics from the dataset from \citet{nayebi_identifying_2020}, which describe the evolution of synaptic weights and activations of layers in various convolutional neural networks trained on multiple tasks with four learning rules. The learning rules utilized in the dataset include Adam (\cite{kingma2014adam}), Stochastic Gradient Descent with Momentum (\cite{sutskever_importance_2013}), Information Alignment (\cite{kunin_two_2020}), and Feedback Alignment (\cite{lillicrap_random_2016}). 
For each analysis, we fixed one task and generated 50 datasets by randomly selecting (without replacement) subsets of trajectories for each learning rule. On each dataset, we fit a HAVOK model with a rank of 45 and lag of 4 to trajectories of these statistics \textit{across} training. We compared each model pairwise, which results in a dissimilarity matrix. After comparing trajectories of these observables across training with \method, we visualize the MDS of the similarity space in Fig. \ref{fig:learning_rules}b. The training dynamics clearly cluster by learning rule under \method. Quantifying the degree of clustering, we could classify the learning rule with 88.5\% test accuracy using only a \textit{linear }classifier and 5-fold cross-validation, whereas the linear SVM classifier in \cite{nayebi_identifying_2020} only reached $\sim 55$\% accuracy (Fig. 2a. in their paper). The test accuracy over a wide range of \method hyperparameters was 81 \% (Fig. \ref{fig:learning_rules}d). 

In the second case, we trained a set of small 3-layer feedforward neural networks (with a hidden size of 100 units) trained to solve a multivariate regression task from \citet{bordelon_influence_2022}. The learning rules in the second set include Gradient Descent, two forms of Feedback Alignment that vary based on the initialization of the random feedback weights (\cite{boopathy_how_2022}), and Hebbian learning (\cite{hebb-organization-of-behavior-1949}). For each learning rule, we trained 50 networks on 1000 epochs with random initializations and randomly generated data. We fit HAVOK models with a delay of 100 and a rank of 500 onto individual networks' activations across learning. As above, we found that the MDS visualization of these networks clustered by learning rule in the pairwise similarity space (Fig. \ref{fig:learning_rules}c). The two types of Feedback Alignment cluster together, suggesting that their dynamics are similar, as might be expected. Once again utilizing a linear classifier, we achieved 89\% test accuracy after assigning both types of FA as the same class, and on average 80\% accuracy over a range of hyperparameters (Fig. \ref{fig:learning_rules}d). These results suggest the \method could be used in a hypothesis-driven fashion to assess learning rules in data recorded from biological circuits.

\section{Discussion}

We demonstrated that our novel metric, \method, can be used to study dynamics in neural networks in numerous scenarios, including the comparison of dynamics between different RNNs, dynamics of a single RNN across training, or dynamics of weight and activity changes in feedforward networks over training. We performed an ablation study in Fig. \ref{fig:schematic}f, which demonstrates that both delay embedding and the DMD were required for \method to succeed. While we applied our metric only to RNNs, we stress that it can be applied to any dynamical system, including diffusion models or generative transformers. Procrustes Analysis over Vector Fields could also be applied to Koopman Embeddings identified via other algorithms, such as an autoencoder (\cite{Lusch_2018}). 

As suggested by Figs.\ref{flipflop}, \ref{galgali}, \ref{ringnetwork} and our theoretical grounding (\ref{equivalency},  \ref{supp:conj}), \method can capture similarity at the level of topology of the dynamics, also known as conjugacy. This can also be recognized from the methodology of \method: the similarity transform is a conjugacy between two linear systems. In recent work, \citet{redman2023equivalent} used the 2-Wasserstein Distance over the Koopman eigenspectrum to compare learning dynamics of fully-connected networks trained to classify handwritten digits. This metric is quite analogous to ours in its theoretical motivation. In Supplementary Information \ref{equivalency} we prove that their metric is a specific case of \method. Thus, \method quantifies how far two dynamical systems are from being conjugate. 

As we demonstrate in Fig. \ref{galgali}, our method could be used to quantitatively compare various hypotheses about the dynamics of a particular biological neural circuit (such as fixed point structure), by calculating similarities of various RNNs or bespoke models to data (\cite{Pagan2022.11.28.518207,schaeffer_reverse-engineering_2020}). This task is typically computationally intensive, relying on assessing the dynamics of noise perturbations in the neural state space (\cite{chaudhuri_intrinsic_2019,galgali_residual_2023}) or computing the locations of fixed points in a task-optimized RNN model that has similar spatial trajectories as the neural data (\cite{mante_context-dependent_2013,CHAISANGMONGKON20171504}). Our method will be useful in the analysis of biological neural systems, where only recorded neural activity is accessible. To our knowledge, \method is the first that can be used to quantitatively assess different dynamical hypotheses. We suggest that this method should be used in conjunction with shape analyses when assessing fits of computational models to data. In future work, we plan to apply \method to experimental data.
% Due to the linearity and composability of our algorithm, much like other similarity metrics, the generalization to second order RSA-style comparison is very straightforward: Because a new DMD matrix can be fit to each condition in a task, we suggest that for each condition, a new $A$ matrix is extracted, then compared pairwise using the Similarity Transform distance used in the paper. This will produce the Representational Dissimilarity matrix, which can then be compared between systems using correlation matrix as in the standard RSA. While we did not explore this application in the paper, it is a straightforward extension that we hope will inspire others to utilize our method. 

%XXX If you don't already include this in discussion, we should; also, are they using dynamical vector fields explicitly for alignment?
% https://www.biorxiv.org/content/10.1101/2022.04.06.487388v1.full.pdf
%Mitchell: yep, that's Karpowicz2022 (below)--not sure about if they're using vector fields, I think they're using LFADS latents (will check)
 % If the fitted HAVOK model is sufficiently accurate, then by \citet{takens} and \citet{koopman1931hamiltonian}, it is possible that the model is topologically equivalent to the original system. Furthermore, \citet{hart_embedding_2020} proved that a sufficiently high dimensional embedding of a nonlinear dynamical system can perfectly predict the next observation with a simple linear readout. 

\method could potentially be used for rapid alignment of Brain-Computer Interfaces (BCI) across days in a subject. Non-stationary sampling of neurons from intracortical electrode shift, neuron death, and other noise plague longitudinal BCI trials, requiring the subject to relearn BCI control daily in a slow and tedious process  (\cite{biran_neuronal_2005,perge_intra-day_2013}). Two recent approaches (\cite{willett_high-performance_2023,degenhart_stabilization_2020}) utilized Procrustes Analysis in their decoding and alignment pipeline. As demonstrated in Figs. \ref{flipflop}, \ref{galgali}, and \ref{ringnetwork}, an alignment method that takes dynamics into account will be useful. \citet{Karpowicz2022} utilized LFADS to align dynamics across days, but \method may provide equal performance with substantially higher data efficiency and less compute due to the method's simplicity. 

In Section \ref{section:learning_rules} we demonstrated how the \method metric space could be used in a hypothesis-driven fashion to compare neurophysiological data recorded across training to artificial neural networks trained with different learning rules. To do so, one would train a variety of neural networks with different learning rules to solve the same task as the animal model. By applying \method pairwise to each network and the neurophysiology data, and identifying which learning rules cluster to the experimental data, we can reject other learning rule hypotheses. \method has multiple advantages over previous methods for this problem: Unsupervised feature generation can improve classification capabilities, and abstraction from geometrymakes the method more generically applicable across systems of different types and initial conditions. 
\vspace{-3mm}
\paragraph{Limitations} It can be challenging to identify the optimal DMD hyperparameters, but it is not computationally intensive to perform a grid sweep. Additionally, we found that the ranks and delays required to achieve our results were relatively small.  On a V100 GPU, a single \method took between 30 seconds and a few minutes, depending on the hyperparameters and size of the dataset.

\paragraph{Acknowledgments} The authors are grateful to Antonio Carlos Costa for bringing Koopman operators to our attention, and to Sarthak Chandra, Raymond Wang, Aran Nayebi, Ho Kyung Sung, Will Redman, Tosif Ahamed, and members of the Fiete lab for helpful discussions and advice.

\bibliography{DMRSA} %

\newpage 
% \documentclass{article}
% \usepackage{algorithm}
% \usepackage{algorithmicx}
% \usepackage{algpseudocode}
% \usepackage{amsmath}
% \usepackage{newclude}

% \usepackage[utf8]{inputenc} % allow utf-8 input
% \usepackage[T1]{fontenc}    % use 8-bit T1 fonts
% \usepackage{hyperref}       % hyperlinks
% \usepackage{url}            % simple URL typesetting
% \usepackage{booktabs}       % professional-quality tables
% \usepackage{amsfonts}       % blackboard math symbols
% \usepackage{nicefrac}       % compact symbols for 1/2, etc.
% \usepackage{microtype}      % microtypography
% \usepackage{xcolor}         % colors
% \usepackage{graphicx}
% \bibliographystyle{plain}
% \usepackage{xspace}
% \usepackage[preprint]{neurips/neurips_2023}
% \input{preamble}

% \newcommand{\method}{DSA\xspace}
% \newcommand{\Tr}{\text{Tr}}

% \title{Supplementary Information}

% \begin{document}

% \maketitle
{\Large \centering Supplementary Information \\}

\section{Dynamical Similarity}
We consider the problem of determining when two linear autonomous dynamical systems on $\fR^n$ 
\begin{equation}
\dt{\x} = \*A\x \hspace{.5cm} \text{and} \hspace{.5cm} \dt{\*y} = \*B\*y 
\end{equation}
may be regarded as equivalent. One natural definition is that for all time $t$, there exists a linear relationship between $\x$ and $\*y$
\begin{equation}
\*y = \*C \*x 
\end{equation}
Differentiating both sides with respect to time and substituting in the dynamics relationship yields an equivalent condition in terms of the system matrices $\*A$ and $\*B$
\begin{equation}
\*y = \*C \*x \iff \*B = \*C \*A \*C^{-1}
\end{equation}
Thus a natural measure of \textit{similarity} between two linear dynamical systems is
\begin{equation}
\text{d}(\*x,\*y) \equiv \min_{\*C}||\*B - \*C \*A \*C^{-1}||
\end{equation}

\subsection{Computing Procrustes Alignment over Vector Fields} \label{supp:optimization}
Unfortunately, this measure of similarity requires solving a nonconvex optimization problem. To simplify, we restrict $\*C$ to be orthogonal, i.e., $\*C^{-1} = \*C^T$ or $\*C \in O(n)$, the orthogonal group. The problem is still nonconvex, but we can take advantage of a convenient parameterization of orthogonal matrices in terms of the \textit{Cayley Transform}. Namely, for an arbitrary skew-symmetric matrix $\*S$, the matrix
\begin{equation}
\*C = (\*I - \*S)(\*I + \*S)^{-1}
\end{equation}
is orthogonal and may readily be computed in deep learning libraries such as PyTorch (e.g., \texttt{C = torch.linalg.solve(I + S, I - S)}).  Using this parametrization, we compute the metric with gradient descent over $\*S$ via the Adam optimizer.  

The Cayley Transform restricts the operation to the special orthogonal group $SO(n)$, which is the group of orthogonal matrices that does not include reflections. We cannot continuously parametrize all of $O(n)$, as it is made up of two disconnected components: $SO(n)$, where for all $C\in SO(n)$,  $\text{det}(C) = 1$, and $O(n) / SO(n)$, where determinants = -1. Note that $SO(n)$ is a subgroup of $O(n)$ while $O(n) / SO(n)$ is not, as $O(n) / SO(n)$ does not have the identity element. To optimize over each with gradient descent, we first state a theorem:

\begin{theorem}
    All elements of $O(n) / SO(n)$ are bijective maps between itself and $SO(n)$.
\end{theorem}
\begin{proof}
    Let $D, P \in O(n) / SO(n)$. Then $PD \in SO(n)$. For this to be true, $PD$ must have two properties: $\text{det}(PD) = 1$, and $(PD)^T(PD) = I$. The first is easily shown: $\text{det}(PD) = \text{det}(P)\text{det}(D) = -1 * -1 = 1$. The second is likewise shown easily: $(PD)^T(PD) = D^TP^TPD = D^TD = I$ by orthogonality. Both $P$ and $D$ have inverses $P^T$ and $D^T$. 

    Conversely, let $D \in SO(n)$. Then we can show by the same logic that $PD \in O(n) / SO(n)$. \end{proof}

From this theorem, we see that we can optimize over $O(n) / SO(n)$ by fixing some matrix in this subset (for example, a permutation matrix that swaps two indices). Then alongside the original formulation of the metric, we can optimize over 

\begin{equation}
    \min_{C \in SO(n)}||B - CP A P^T C^T||_F
\end{equation}

and take the minimum of the two values to be our final score.

\section{The minimum similarity transform is a metric} \label{supp: metric}

Let A and B each be matrices in $\mathcal{R}^{n \times n}$. We wish to show that 

\begin{equation}
    d(A,B) = \min_{C \in O(n)} ||A-CBC^{-1}||_F
\end{equation} is a proper metric--that is, the following three properties hold:
\begin{itemize}
    \item Zero: $d(A,B) = 0 \iff A = B$ 
    \item Symmetry: $d(A,B) = d(B,A)$ 
    \item Triangle Inequality: $d(A,C) \leq d(A,B) + d(B,C)$
\end{itemize}

\begin{proof}
According to \cite{williams_generalized_2022}, only two things are needed. First, we must show that 
\begin{equation}
    g(A,B) = ||A - B||_F
\end{equation} is a metric, which is trivially true. Then we must show that the similarity transform is an isometry: 
\begin{equation}
    g(T(A),T(B)) = g(A,B) 
\end{equation}
where $T(A)$ is a map that in our case is the similarity transform. 
\begin{equation}
    g(T(A),T(B)) = ||CAC^{-1} - CBC^{-1}||_F = ||C(A-B)C^{-1}||_F
\end{equation}
Using the identity that the Frobenius norm of a matrix is equivalent to the trace of the inner product with itself,
\begin{equation}
    g(T(A),T(B)) = \Tr([C(A-B)C^{-1}]^T[C(A-B)C^{-1}]) 
\end{equation}
\begin{equation}
    = \Tr(C(A-B)^TC^{-1}C(A-B)C^{-1})
\end{equation}
\begin{equation}
= \Tr(C(A-B)^T(A-B)C^{-1}) = \Tr(C^{-1}C(A-B)^T(A-B)) 
\end{equation}
\begin{equation}
    = \Tr((A-B)^T(A-B)) = ||A-B||_F = g(A,B)
\end{equation}
With the last steps using the cyclic property of the trace ($\Tr$) and the fact that C is an orthonormal matrix.
\end{proof}

\section{The minimizer of the euclidean distance also minimizes angular distance}
\begin{proof} Let $C^*$ be the minimizer of $||A-CBC^{-1}||_F$. Then we can rewrite this as above, using the trace identity:
\begin{equation}
    ||A-CBC^{-1}|| = \Tr((A-CBC^{-1})^T(A-CBC^{-1}))
\end{equation}
which can be expanded into:
\[= \Tr(A^TA) - 2\Tr(A^TCBC^{-1}) + \Tr(B^TB)\]
\[ = \Tr(A^TA) - 2<A,CBC^{-1}> + \Tr(B^TB)\]

By minimizing the Frobenius norm, $C^*$ must maximize the inner product $<A,CBC^{-1}>$. The angular distance is defined as 
\begin{equation}
\arccos{\frac{<A,CBC^{-1}>}{||A|| ||B||}}
\end{equation}
Due to the arccos function being montonically decreasing, maximizing the matrix inner product therefore minimizes the angular distance. \end{proof}

\begin{figure}[H] 
    \centering
    \includegraphics[width=\linewidth]{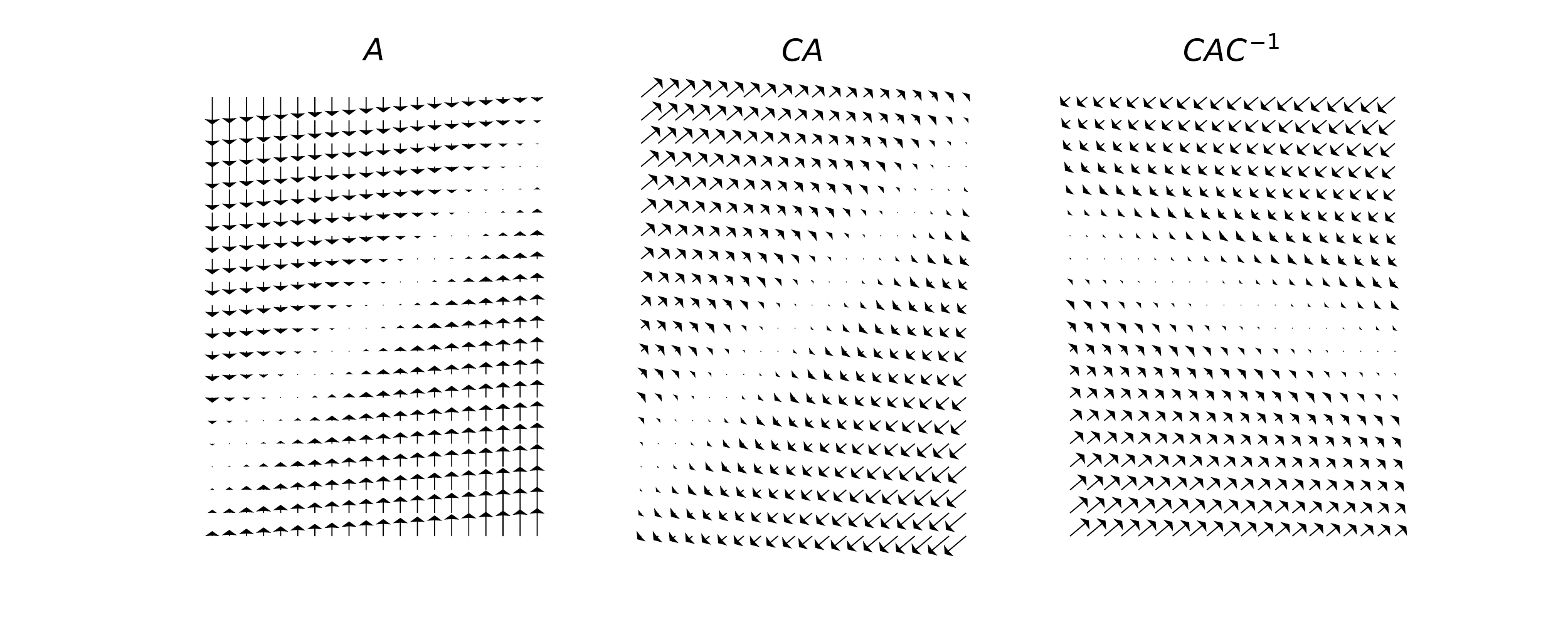}
    \caption{Vector Fields for a linear system $A$ and two transformations. Here, $C$ is an orthogonal matrix. The size of the arrow represents the magnitude of the vector, so the small dots are zeros, indicating the line attractor.}
\end{figure}

\section{Vector Fields Under Orthogonal Transformation} \label{supp:orthog}
Here we demonstrate how Vector Fields transform. Given some linear system $\dt{x} = Ax$, the matrix $A$ contains all the information about the vector field (Fig. 6,left).  The transformation $\*C \*A_y$ simply rotates vectors in place, which destroys dynamic structure (for example, it could turn a stable fixed point into an unstable one). However, $\*C\*A_y\*C^{-1}$ rotates their positions as well, thereby preserving the vector field structure and satisfying the same goal as $\*C \*Y$ applied to a data matrix. 

Applying a single orthogonal matrix $C$ to the left (or right) of $A$ to yield the system $\dt{x}=CAx$ produces the plot in  Fig. 6 middle. This is clearly a different dynamical system, as all the vectors now point away from the line attractor attractor, thereby making the system unstable instead of stable. However, applying the similarity transform $CAC^{-1}$ preserves the dynamical structure of the linear system by rotating both the vectors and their positions on the field (Fig 6, right).

\section{Equivalency to 2-Wasserstein Distance} \label{equivalency}
Here we prove a relationship between DSA and the 2-Wasserstein distance $W_2(X,Y)$. Because the equivalence of the eigenvalues in $A_x$ and $A_y$ is sufficient to show conjugacy of the two underlying systems (if they have point spectra, next section), the Wasserstein distance is also a useful metric to measure distance between two dynamical systems. This was utilized in \citet{redman2023equivalent} to measure dynamical similarities between different learning algorithms. 
\begin{theorem}
DSA is equivalent to the 2-Wasserstein Distance between the eigenvalues of $A_x$ and $A_y$ when these matrices are normal.
\end{theorem}
\begin{proof} 
\begin{equation}
    DSA(A_x,A_y) = \min_{C \in O(n)}\left| A_x - CA_yC^T \right|_F
\end{equation}
\begin{equation}
W_2(X,Y) = \inf_\pi \left(\frac{1}{n}\sum_{i=1}^n \left|X_{ii} - Y_{\pi(ii)}\right|_2 \right)^{1/2} = \min_{P \in \Pi(n)}\left|X - PYP^{-1}\right|_F
\end{equation}
Where $X$ and $Y$ are sets represented as diagonal matrices. $\Pi(n)$ is the group of permutation matrices, a subgroup of the orthogonal group $O(n)$.
When $A_x$ and $A_y$ are normal, they can be eigendecomposed as $A_x = V_x \Lambda V_x^T$ and $A_y = V_y \Delta V_y^T$. $\Lambda$ and $\Delta$ are diagonal matrices of eigenvalues. 
By the group properties of $O(n)$, $C$ can be written as the composition of other elements in the group. 
Let $C = V_xPV_y^T$ where $P \in O(n)$. Then $DSA$ is equivalent to 
\begin{equation}
    DSA(A_x,A_y) = \min_{P \in O(n)}\left| V_x\Lambda V_x^T - V_x P\Delta P^T V_x^T \right|_F = \min_{P \in O(n)}\left| \Lambda - P\Delta P^T \right|_F 
\end{equation}
Now we wish to show that the minimum occurs when P is a permutation matrix. Using the trace identity of the Frobenius norm, we can rewrite this as 
\begin{equation}
    DSA(\Lambda, \Delta) = \min_{P \in O(n)} \sqrt{\text{Tr}((\Lambda - P\Delta P^T)^T(\Lambda - P\Delta P^T))}
\end{equation}
which is minimized at the maximum of $\text{Tr}(\Lambda^TP\Delta P^T)$. In summation notation,
\begin{equation}
    \max_{P \in O(n)}\text{Tr}(\Lambda^TP\Delta P^T) = \max_{\sum_{j}p_{ij}^2 = 1} \sum_i \Lambda_i \sum_j p_{ij}^2 \Delta_j = \max_{P \in O(n)} \text{Tr}(\Lambda P \circ P \Delta)
\end{equation}
$P \circ P$ is a doubly stochastic matrix, the set of which is a convex polytope whose vertices are permutation matrices (Birkhoff-von Neumann). 
$\text{Tr}(\Lambda P \circ P \Delta)$  is affine in $P \circ P$, which means its critical points are those vertices. Thus, $P \circ P$ and $P$ itself is a permutation matrix, $P \in \Pi(n)$.
\end{proof}

\section{Conjugacy and the Eigenspectra of the Koopman Operator} \label{supp:conj}
In our experiments, we demonstrated that \method can identify conjugacy between different dynamical systems. But is this generally true, and do we have any theoretical guarantees on this? Here we recapitulate results from Koopman theory to show that \method indeed has the properties necessary to identify conjugate dynamics. 

First, recall that HAVOK and other DMD methods seek to identify a finite approximation of the Koopman operator from data. The Koopman operator is defined as the shift map of observables, $U_f^t\theta(x_0) = \theta (f^t(x_0))$, where our dynamical system is defined by the flow function $f^t$, with a time parameter that takes the state from $x_0$ to $x_t$. $\theta : X \rightarrow \mathbb{C}$ is function called an observable. Second, note that a similarity transform $CAC^{-1}$ preserves the eigenvalues and multiplicities of the matrix $A$. As seen above in Section \ref{equivalency}, under some conditions \method can be viewed as comparing the eigenvalues of the DMD matrices $A_x$ and $A_y$. Assuming that we are in the regime such that the DMDs converges to the Koopman approximation, the following result from \citet{budisic_applied_2012} implies that \method will report zero if the systems are conjugate. 
\begin{theorem}[Spectral equivalence of topologically conjugate systems \cite{budisic_applied_2012}]
Let $S : M \rightarrow M$ and $T : \rightarrow N \rightarrow N$ be maps that define two topologically conjugate dynamical systems over different manifolds $M$ and $N$. Let there exist a homeomorphism $h : N \rightarrow M$ such that $S \circ h = h \circ T$. If $\phi$ is an eigenfunction of the Koopman operator of $S$, $U_s$ (i.e. $U_s \phi = \lambda \phi$), then $\phi \circ h$ is an eigenfunction of the Koopman operator of $T$, $U_t$ also with eigenvalue $\lambda$.
\end{theorem}
\begin{proof}
    Define $x \in M$ and let $y \in N$ such that $x = h(y)$. Then
 
      $$U_S\phi(x) = \phi S(x) = \lambda \phi(x)$$
      $$\phi \circ (S \circ h)(y) = \lambda (\phi \circ h)(y)$$
      $$\phi \circ (h \circ T)(y) = \lambda (\phi \circ h)(y)$$
      $$U_T (\phi \circ h)(y) = \lambda (\phi \circ h)(y)$$ \end{proof}

We would also like to show the converse: if two Koopman operators have the same eigenspectra, their underlying systems are conjugate. We briefly sketch a proof here, which is captured in this commutative diagram. Here, $f$ and $g$ represent our nonlinear systems, with Koopman operators $U^f$ and $U^g$. $A_f$ and $A_g$ represent the valid linearizations: 

\[
\begin{tikzcd}
f \arrow[d, "U^f", leftrightarrow]
& g \arrow[d, "U^g",leftrightarrow] \\
A_f \arrow[r, "C^{-1}(\cdot)C", leftrightarrow]
& A_g
\end{tikzcd}
\]

As seen in \citet{meziclinearization}, we can construct linearizations of nonlinear dynamical systems from Koopman eigenfunctions, that hold over an entire attractor basin. This corresponds to the vertical arrows in the diagram. With these in hand, we can use a basic result in linear systems theory which also inspired the metric in \method: Two linear dynamical systems are conjugate if their eigenvalues are equivalent, and the conjugate mapping is linear. Because these linear systems are one-to-one with the given nonlinear systems in a particular attractor basin, the diagram holds. Because each arrow is a conjugate mapping, the composition of three conjugacies is a conjugacy and thus the $f$ and $g$ are conjugate themselves. 

% Here, $f$ and $g$ are two dynamical systems: $\dt{x} = f(x)$ and $\dt{y} = g(y)$. $\phi$ is a homeomorphism, and K is the Koopman operator. Then our matrices $A_x$ and $A_y$ are therefore related with the similarity transform.

\section{\method pseudocode}

\algrenewcommand\algorithmicrequire{\textbf{Input:}}
\algrenewcommand\algorithmicensure{\textbf{Output:}}
\begin{algorithm}[H]
\caption{Dynamic Similarity Analysis}\label{alg:cap}
\begin{algorithmic}
\Require $X_1,X_2 \in \mathbb{R}^{n \times t \times d}$, $p$, $r$
\Ensure Similarity transform distance $d$ between the two dynamics matrices
\Procedure{DelayEmbedding}{$X$, $p$}
    \State Initialize $H \in \mathbb{R}^{n \times (t-p) \times p*d}$
    \State $H = \begin{bmatrix} X[:, 1:p, :] \\ X[:, 2:p+1, :] \\ \cdots \\ X[:, t-p+1:t, :] \end{bmatrix}$ reshaped to $\mathbb{R}^{n \times (t-p) \times p*d}$
    \State \Return H
\EndProcedure

\Procedure{DMD}{$H$,$r$}
\State Initialize $A \in \mathbb{R}^{r \times r}$
\State $H' = H[:, 1:]$
\State $H = H[:,:-1]$
\State Decompose $H = U\Sigma V^T$, and $H' = U'\Sigma'V'^T$
\State Reduce Rank: $V = V[:,:r]$, and $V' = V'[:,:r]$
\State Compute $A$ such that $V' = AV$ (OLS or Ridge Regression, depending on regularization)
\State \Return A
\EndProcedure

\State $H_1 =$  \Call{DelayEmbedding}{$X_1$, $p$} 
\State $H_2 = $ \Call{DelayEmbedding}{$X_2$,p}

\State $A_1 =$ \Call{DMD}{$H_1$,r}
\State $A_2 = $ \Call{DMD}{$H_2$,r}

\State $d = \min_{\*C \in \*O(r)}||\*A_1 - \*C \*A_2 \*C^{-1}|| $

\end{algorithmic}
\end{algorithm}
\vspace{-0.5cm} 

\section{Eigen-Time Delay Coordinates is equivalent to PCA whitening.} \label{supp:whiten}
In the Eigen-Time Delay Coordinate formulation of HAVOK, we compute the DMD over the right singular vectors of the SVD of H, $H = U\Sigma V^T$. PCA whitening normalizes the covariance matrix of the (already centered) data $H$:
\begin{equation}
    HH^T = U\Sigma^2U^{-1}
\end{equation}
PCA whitening is equivalent to $H' \leftarrow \Sigma^{-1}U^{-1}H$. PCA whitening applied to the SVD of H is written as
\begin{equation}
    H' \leftarrow \Sigma^{-1}U^{-1}U^\Sigma V^T = V^T
\end{equation}
which is evidently the eigen-time delay formulation.

\section{FlipFlop task MDS clustering by activation} \label{supp:mds_act}
Here, we plotted the same MDS scatterplots as in Fig. 2 in the main text. The only difference here is that the networks are colored by activation function instead of architecture. Evidently, the same conclusion holds here as in Fig. 2.
\begin{figure}[H]
    \centering
    \includegraphics[width=\linewidth]{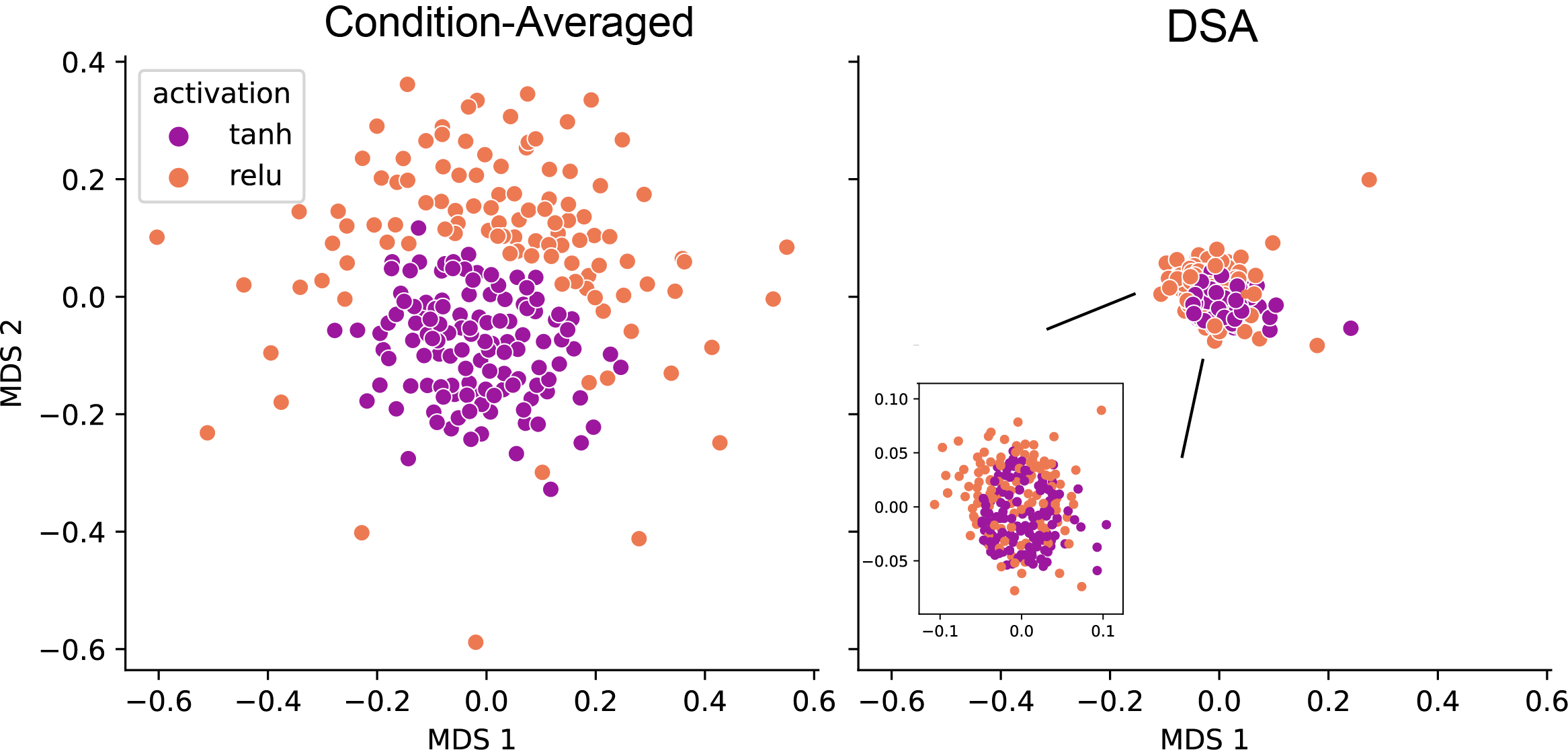}
    \caption{\textbf{Same dynamics, different shape, only identified as similar with \method }. Dots indicate a single trained network. Analysis of RNNs trained on the 3-bit flipflop task. \textbf{b.} MDS Projection of the Dissimilarity Matrix computed across condition-averaged hidden states between each RNN (activation function indicated by color). In this view of the data, RNNs cluster by activation function. \textbf{c.} MDS Embedding of the dissimilarity matrix generated from \method of the same set of RNNs as in b. Here, RNNs do not cluster by activation function. }
\end{figure}

\section{Flip-Flop task HAVOK grid sweep} \label{supp:heatmap}
We fit HAVOK models with a range of ranks and delays, and assessed their fits in three different ways: $R^2$, MSE, and Pearson correlation between the predicted state from HAVOK and the true data, across 10 different models. Note that we did not exhaustively test all trained models and a large range of ranks and delays for the sake of efficiency. Here we see that across all metrics, steadily increasing rank and number of delays (listed as lag) improves the performance of the model. Across all metrics, there appears to be a Pareto Frontier which implies that the specific choice of rank and lag is not important. Furthermore, the metrics appear to saturate relatively quickly. 

\begin{figure}[H]
    \centering
    \includegraphics[width=\linewidth]{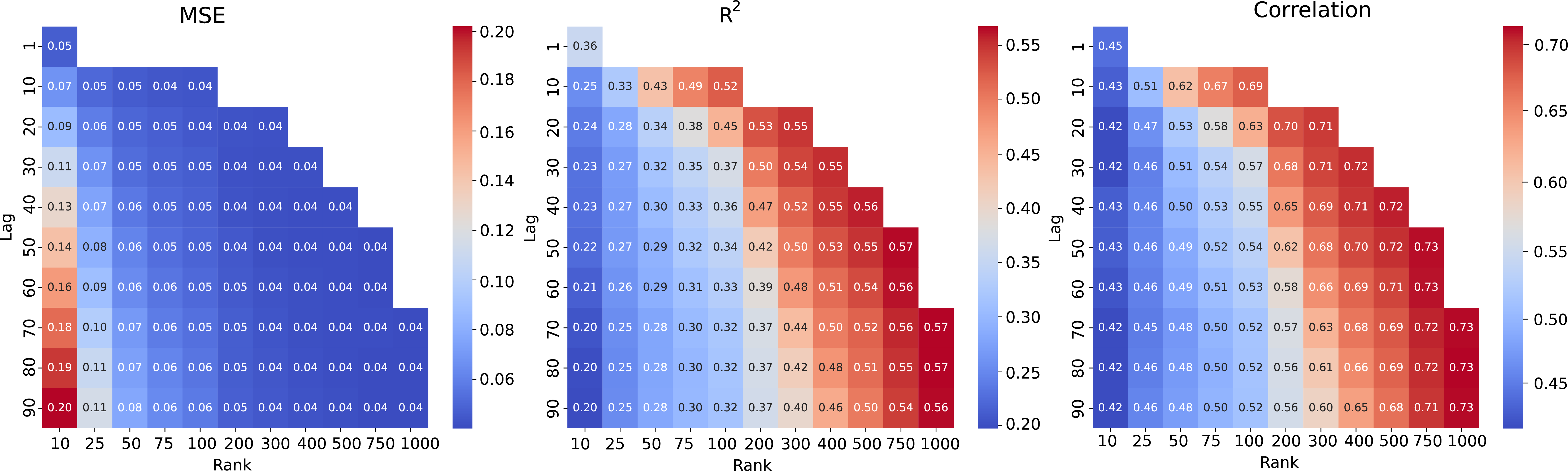}
    \caption{Mean HAVOK fits for three different scores across a variety of delays and ranks. The unfilled elements occurs when the rank is larger than the number of lags times the delays. Each model was computed over 15 Principal Components of the data, which had 128 trials with 500 timesteps each. }
\end{figure}

\section{Condition-Averaged Aligned Dynamics Systems Equations} \label{resid_eqn}
Here we describe the process we used to sample from the three dynamical systems with different attractors, and adversarially optimize input so that the condition averages are indistinguishable. This describes the systems tested in Figure 3 in the main text. 
\paragraph{Bistable System}
\begin{equation}
    \dt{x}_1 = ax_1^3 +bx_1 + u + \epsilon_2 
\end{equation}

\begin{equation}
    \dt{x}_2 = cx_2 + u + \epsilon_2
\end{equation}
We randomly sample parameters for the system once  from: $a \sim U(-5,-3) \text{ } b \sim U(4,7) \text{  } c \sim U(-4,-2)$  and noise at each time step sampled i.i.d. $ \epsilon \sim N(0,0.05)$ We update the system with an Euler update with $dt = 0.01$. $u$ is defined as a constant in $\{-0.1,0.1\}$ to drive the system to either of the stable fixed points.

\paragraph{Line Attractor}
We define the eigenvectors and eigenvalues to get the linear system:
\begin{equation}
    \Lambda = \begin{pmatrix}
        -1 & 0 \\ 
        0 & 0
    \end{pmatrix} \quad   V = \begin{pmatrix}
        1 & 1 \\ 
        1 & 0
    \end{pmatrix} \quad     A = V \Lambda V^{-1}
\end{equation}
\begin{equation}
    \dt{x} = A x + \hat{u}
\end{equation}

\paragraph{Point Attractor}
\begin{equation}
    A = \begin{pmatrix}
        -0.5 & 0 \\ 
        0 & -1
    \end{pmatrix}
\end{equation}
\begin{equation}
    \dt{x} =  Ax + \hat{u}
\end{equation}

\paragraph{Adversarial optimization scheme to align condition averages}
We apply a simple feedback linearization scheme to control the condition averages of the line and point attractor. We simulate batches of trajectories simultaneously in order to calculate condition averages for the optimization scheme. First, we identify the condition averages $\bar{y}$ from the bistable system, which are previously calculated. At each step of the latter two systems, we calculate the condition averages $\bar{x}$. Then we define $\hat{u}$, the input that aligns the condition averages between systems:

\begin{equation}
    \hat{u} = -A\bar{x}_t + \frac{1}{\alpha}(\bar{y}_{t+1} -\bar{x}_t)
\end{equation}
Thus, when we apply the Euler update with time step $\alpha$, our system is defined by
\begin{equation}
    x_{t+1} = x_t + \alpha(Ax_t - A\bar{x}_t) + \bar{y}_{t+1} - \bar{x}_t
\end{equation}
Which makes our condition-averaged dynamics of the line and point attractor system
\begin{equation}
    \bar{x}_{t+1} = \bar{y}_{t+1}
\end{equation}
\section{Ring Attractor Dynamics} \label{supp:ring}
We implemented the ring attractor defined in \cite{wang2022}, specified by the following equation:
\begin{equation}
    \tau \frac{ds}{dt} = - s + W^T\phi (s) + A \pm \gamma b(t) + \xi (x,t)
\end{equation}
Here, $s$ describes the synaptic activation along the ring, and $\phi$ is the neural transfer function that converts these into firing rates: $r = \phi(s)$. In our simulations, $\phi$ was the ReLU function. To deform the geometry while preserving the ring topology, we applied the sigmoid transform in the main text to the extracted synaptic activations $s$ after simulation. To stabilize the dynamics of the ring for path integration, there are in fact two rings, each with their connectivity profile slightly offset, and with the $\pm$ term separately $+$ for one ring and $-$ for the other. $A$ describes the resting baseline input to the neurons, whereas $b$ describes the driving input that the network seeks to integrate, and $\gamma$ is a coupling term. $\tau$ describes the time constant, and $\xi$ describes the activity noise term. $W$ of course describes the weight matrix that has local excitatory connectivity and global inhibitory connectivity between neurons in the ring. 

\section{Line Attractor Deformation}

We use the same line attractor matrix as in section \ref{resid_eqn}:
\begin{equation}
    \Lambda = \begin{pmatrix}
        -1 & 0 \\ 
        0 & 0
    \end{pmatrix} \quad   V = \begin{pmatrix}
        1 & 1 \\ 
        1 & 0
    \end{pmatrix} \quad     A = V \Lambda V^{-1}
\end{equation}

\begin{figure}[H]
    \centering
    \includegraphics[width=\linewidth]{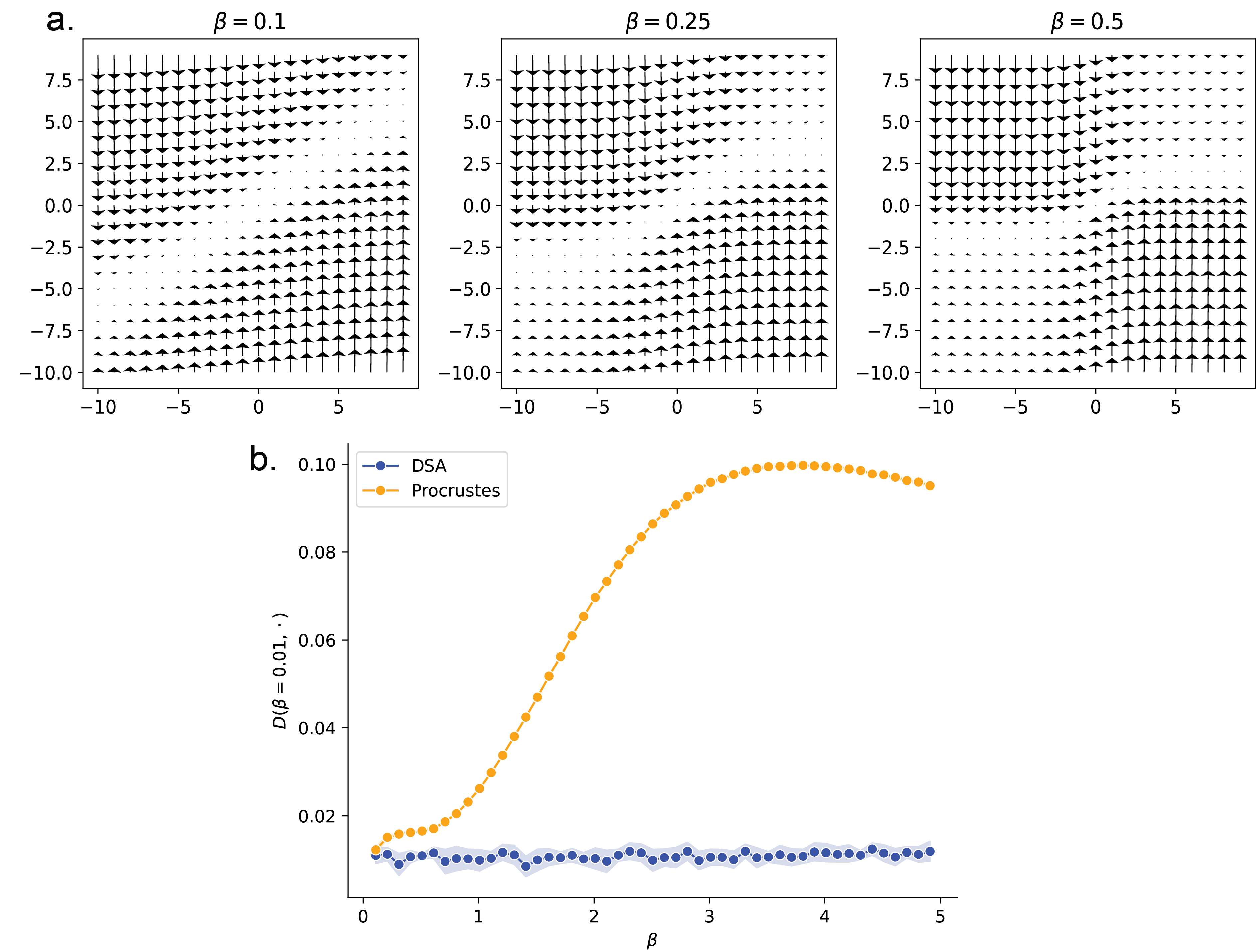}
    \caption{\textbf{\method is invariant under geometric deformations when applied to a line attractor}. \textbf{a}. Demonstration of how the vector field transforms under $\beta$. Here, the line attractor becomes more s-shaped at higher $\beta$ values. \textbf{b}. Procrustes and \method similarity to the smallest tested $\beta$ (0.1) at various values. Shaded region indicates standard error of the mean over 10 simulations.}
    \label{lineattractor_tanh}
\end{figure}

However, the dynamics are slightly different, as we introduce a nonlinear activation function: 
\begin{equation}
    \dt{x} = A\tanh{(\beta x)} + u + \epsilon
\end{equation}
Where $u = \{1,-1\}$ and is constant across both dimensions to drive the system along the attractor in different directions. Here, $\beta$ is a scalar that affects the curvature of the attractor, as demonstrated by the vector fields of this system at three example values (Fig. \ref{lineattractor_tanh}a). We applied the same analysis as in Fig.4 in the main text and found the exact same results--Procrustes distance increases in magnitude as the two $\beta$ values being compared increases, but \method is invariant. This adds to the evidence that \method captures the topology of a dynamical system.  

\end{document}